\documentclass[fleqn,usenatbib]{mnras}

\usepackage{newtxtext,newtxmath}
\usepackage[T1]{fontenc}
\usepackage{ae,aecompl}
\usepackage{relsize}

\usepackage{graphicx}	
\usepackage{amsmath}	
\usepackage{amssymb}	

\title[Modified high-order Hermite integrators]{Modified Hermite Integrators of Arbitrary Order}

\author[A. J. Dittmann]{
Alexander J. Dittmann$^{1}$\thanks{E-mail: \href{mailto:dittman@astro.umd.edu}{dittmann@astro.umd.edu}}
\\
$^{1}${Department of Astronomy, University of Maryland, College Park, MD 20742-2421}\\
}

\date{Accepted 2020 June 1. Received 2020 May 22; in original form 2020 April 16}

\pubyear{2020}

\begin{document}
\label{firstpage}
\pagerange{\pageref{firstpage}--\pageref{lastpage}}
\maketitle

\begin{abstract}
We present a family of modified Hermite integrators of arbitrary order possessing superior behaviour for the integration of Keplerian and near-Keplerian orbits. After recounting the derivation of Hermite N-body integrators of arbitrary order, we derive a corrector expression that minimises integrated errors in the argument of periapsis for any such integrator. In addition to providing an alternate derivation of the modified corrector for the 4th-order Hermite integrator, we focus on improved correctors for the 6th- and 8th-order integrators previously featured in the literature. We present a set of numerical examples and find that the higher-order schemes improve performance, even when considering their slightly higher cost in floating point operations. The algorithms presented herein hold promise for systems dominated by central potentials, such as planetary systems and the centres of galaxies. Existing Hermite integrators of any order can be modified to use the expressions presented here with minimal effort. Accordingly the schemes presented herein can be easily implemented on massively parallel architectures. 
\end{abstract}



\section{Introduction}

The 4th-order Hermite algorithm is very popular for studying stellar dynamics \citep[e.g.][]{{1991ApJ...369..200M},{1999PASP..111.1333A},{2016MNRAS.463.2109R},{2016MNRAS.458.1450W}}, as well as planetesimals systems \citep[e.g][]{{2004PASJ...56..861K},{2007ApJ...669.1316T},{2007ApJ...671.2082K},{2010ApJ...714L..21K}}. Hermite integration 
schemes of order 6 and 8 have been derived by \citet{2008NewA...13..498N}, although these methods have sub-optimal truncation error for orbits that are nearly Keplerian. Errors in the eccentricity vector over time can be reduced significantly at the cost of a few floating point operations per particle each time step. 
When used with a time-symmetric time step, the methods presented herein evolve systems such that errors in energy and eccentricity remain bounded and errors in the argument of periapsis are minimised,
making them suitable for the study of planetesimal discs, stellar discs about supermassive black holes (SMBHs), and planetary systems. In particular, this algorithm was motivated by models of stellar migration through accretion discs \citep{{2019arXiv191208218T},{2020MNRAS.493.3732D},{2020MNRAS.494.1203M}}. The high order of these methods, as well as their ability to readily support time-symmetric variable time steps make them applicable to systems that are largely Keplerian as well as those with frequent close encounters. 

The methods derived herein offer the most benefit for Hermite implementations that use multiple predictor-corrector iterations, making the methods more symmetric in time. Additionally, higher-order Hermite schemes
can be
unstable unless used with multiple predictor-corrector iterations \citep{2015nbdywrk}. 
High-order Hermite N-body algorithms have been preferred because their larger time step sizes outweigh their cost in floating point operations, and because of their improved treatment of interactions between particles with large mass ratios \citep{2015ComAC...2....8B}.
The present work extends these benefits to simulations of planetary and protoplanetary systems along the lines of the methods presented in \citet{2004PASJ...56..861K}.

We have derived corrector expressions that minimise errors when integrating Keplerian orbits, which we refer to as modified Hermite integrators. This derivation would not be possible without the work presented in \citet{2016RIKEN.K.N} at a lab seminar at RIKEN, which covers the derivation of Hermite methods of arbitrary order but does not alter such methods to properly treat Keplerian orbits. Because this material is not widely available in the literature and it is necessary to understand how to modify truncation errors to preserve other conserved quantities, we recount relevant portions of the derivations presented in \citet{2016RIKEN.K.N} before extending them. Afterwards we modify the 6th- and 8th-order Hermite algorithms presented in \citet{2008NewA...13..498N}, and verify that they behave as expected through numerical experiments. Afterwards, we comment on the applicability and future prospects for high-order time-symmetric Hermite N-body integrators. 

\section{Derivation of High-Order Integrators}
The first portion of this section is largely drawn from the work of Keigo Nitadori presented in \citet{2016RIKEN.K.N}. Derivations of this form can be found in his seminar, but basic 6th- and 8th-order methods were presented in \citet{2008NewA...13..498N}, and 10th- and 12th-order methods were discussed in \citet{2015nbdywrk} . Afterwards, we derive a modified corrector expression for any order that optimises behaviour for the Kepler problem.

Consider a vector $\textbf{f}$ and its corresponding derivatives $\textbf{f}^{(n)}=\frac{d^n}{dt^n}\textbf{f}(t)$ up to the $p$-th, with $0\leq n\leq p$. We adjust the dimension of the derivatives by a step size $ h\equiv\Delta t/2$, such that
\begin{equation}
\textbf{F}(t) = 
\begin{bmatrix}
\textbf{f}(t)\\
h\textbf{f}^{(1)}(t)\\
h^2\textbf{f}^{(2)}(t)/2!\\
\vdots \\
h^p\textbf{f}^{(p)}(t)/p!\\
\end{bmatrix}
\end{equation}
obeys the differential equation
\begin{equation}
\frac{d}{dt}\textbf{F}(t)=\frac{1}{h}
\begin{bmatrix}
0 & 1 & 0 & \cdots & 0 \\
0 & 0 & 2 & \cdots & 0 \\
0 & 0 & 0 & \cdots & \vdots \\
\vdots & \vdots & \vdots & \ddots & p \\
0 & 0 & 0 & \cdots & 0
\end{bmatrix}\textbf{F}(t)
\end{equation}
which has a formal solution at time $t+h$ given by multiplication by an upper triangular Pascal matrix \footnote{\href{https://en.wikipedia.org/wiki/Pascal_matrix}{https://en.wikipedia.org/wiki/Pascal\_matrix}} 
\begin{equation}
\textbf{F}(t+h)=
\begin{bmatrix}

\left(\begin{smallmatrix} 0\\0 \end{smallmatrix}\right) & \left(\begin{smallmatrix} 1\\0 \end{smallmatrix}\right) & \left(\begin{smallmatrix} 2\\0 \end{smallmatrix}\right) & \cdots & \left(\begin{smallmatrix} p\\0 \end{smallmatrix}\right) \\
0 & \left(\begin{smallmatrix} 1\\1 \end{smallmatrix}\right) & \left(\begin{smallmatrix} 2\\1 \end{smallmatrix}\right) & \cdots & \left(\begin{smallmatrix} p\\1 \end{smallmatrix}\right) \\
0 & 0 & \left(\begin{smallmatrix} 2\\2 \end{smallmatrix}\right) & \cdots & \left(\begin{smallmatrix} p\\2 \end{smallmatrix}\right) \\
\vdots & \vdots & \vdots & \ddots & \vdots \\
0 & 0 & 0 & \cdots & \left(\begin{smallmatrix} p\\p \end{smallmatrix}\right) \\ 
\end{bmatrix}\textbf{F}(t)
\end{equation}
\citep{2016RIKEN.K.N}, where the parentheses in each matrix element represent binomial coefficients.
Then, we define
\begin{equation} 
\textbf{F}_n = \frac{h^n}{n!}\textbf{f}^{(n)}(t)
\end{equation}
\begin{equation}
\textbf{F}^\pm_n=\frac{1}{2}\frac{h^n}{n!}\left(\textbf{f}^{(n)}(t+h)\pm \textbf{f}^{(n)}(t-h)\right).
\end{equation}
When updating velocity, we take $\textbf{f}=\textbf{a}$, and for position we take $\textbf{f}=\textbf{v}$, where $\textbf{a}$ is the acceleration and $\textbf{v}$ is the velocity.
We can then update the solution using the even terms, which survive by parity:
\begin{equation}
\Delta \textbf{v} = \int_{-h}^h \textbf{f}(t)dt \approx\left( \textbf{F}_0 + \frac{1}{3}\textbf{F}_2 + ... + \frac{1}{2p+1}\textbf{F}_{2p} \right)\Delta t,
\end{equation}
where $2(p+1)$ is the order of the integrator \citep{2016RIKEN.K.N}. 
When updating the position, we have calculated enough derivatives of the acceleration to add another term, 
\begin{equation}
\Delta \textbf{x} = \int_{-h}^h \textbf{f}(t)dt \approx \left(\textbf{F}_0 + ... + \frac{1}{2p+1}\textbf{F}_{2p} + \frac{\beta}{2p+3}\textbf{F}_{2p+2}\right)\Delta t,
\end{equation}
where $\beta$ in the final term controls the truncation error in position. The additional term has no influence on the overall order of the integration scheme, but can improve conservation of other quantities such as the eccentricity vector of an orbit. Note that each $\textbf{F}_n$ is different for $\Delta\textbf{x}$ than $\Delta\textbf{v}$, and that the velocity must be calculated first so that it can be used in the position calculation. 

We can calculate $\Delta \textbf{v}$ and $\Delta \textbf{x}$ by solving the system of equations 
\begin{equation}
\begin{bmatrix}
\textbf{F}^+_0 \\
\textbf{F}^-_1 \\
\textbf{F}^+_2 \\
\textbf{F}^-_3 \\
\vdots \\
\end{bmatrix}
=
\begin{bmatrix}
\left(\begin{smallmatrix} 0\\0 \end{smallmatrix}\right) &\left(\begin{smallmatrix} 2\\0 \end{smallmatrix}\right) &
\left(\begin{smallmatrix} 4\\0 \end{smallmatrix}\right) &\left(\begin{smallmatrix} 6\\0 \end{smallmatrix}\right) & \cdots \\
0 &\left(\begin{smallmatrix} 2\\1 \end{smallmatrix}\right) &
\left(\begin{smallmatrix} 4\\1 \end{smallmatrix}\right) &\left(\begin{smallmatrix} 6\\1 \end{smallmatrix}\right) & \cdots \\
0 &\left(\begin{smallmatrix} 2\\2 \end{smallmatrix}\right) &
\left(\begin{smallmatrix} 4\\2 \end{smallmatrix}\right) &\left(\begin{smallmatrix} 6\\2 \end{smallmatrix}\right) & \cdots \\
0 & 0 & \left(\begin{smallmatrix} 4\\3 \end{smallmatrix}\right) &\left(\begin{smallmatrix} 6\\3 \end{smallmatrix}\right) & \cdots \\
\vdots & \vdots & \vdots & \vdots & \ddots \\
\end{bmatrix}
\begin{bmatrix}
\textbf{F}_0 \\
\textbf{F}_2 \\
\textbf{F}_4 \\
\textbf{F}_6 \\
\vdots \\
\end{bmatrix}
=\underline{A}
\begin{bmatrix}
\textbf{F}_0 \\
\textbf{F}_2 \\
\textbf{F}_4 \\
\textbf{F}_6 \\
\vdots \\
\end{bmatrix},
\end{equation}
where $\underline{A}$ is a matrix formed from the even columns of the upper triangular Pascal matrix \citep{2016RIKEN.K.N}. From this form, it is clear that we can construct 
an integration scheme of order $2(p+1)$ by calculating $p$ derivatives of acceleration. Otherwise put, we can use linear combinations of lower-order even and odd derivatives to evaluate higher-order even derivatives. 

\subsection{Optimal coefficients for the Kepler problem}
In this section, we present a derivation of the optimal value of $\beta$ for the Kepler problem. It is useful to present this derivation, to better conserve the argument of periapsis, because one can straightforwardly modify this procedure to conserve other quantities. In particular, we point out that time symmetric Hermite methods can be applied to systems with energy-conserving velocity-dependent effects, such as Coriolis acceleration and the Lorentz force, where one may wish to conserve other quantities. 

For an integrator of order $2(p+1)$ the leading order term in the truncation error for $\Delta \textbf{v}$ comes from our omission of the term 
$\frac{1}{2p+3}\textbf{F}_{2p+2}$ as well as neglecting the corresponding column in the Pascal matrix. When calculating $\Delta \textbf{x}$ up to the $\textbf{F}_{2p+2}$ term, we retain the appropriate columns of the Pascal matrix, and are left with leading truncation error resulting from only our choice of $\beta$. 
The coefficient for the $p$-th term of an order $2(p+1)$ Hermite integrator is
\begin{equation}
\frac{1}{(-2)^p}\frac{(2p)!!}{(2p+1)!!}
\end{equation}
\citep{2016RIKEN.K.N}.
Thus, the truncation error in velocity ($\delta \textbf{v}$) for an order $2(p+1)$ Hermite integrator is 
\begin{equation}
\delta \textbf{v} = \frac{-1}{(-2)^{p+1}}\frac{(2p+2)!!}{(2p+3)!!}\textbf{F}^\pm_{p+1}.
\end{equation}
The truncation error in \textbf{x} ($\delta \textbf{x}$) only comes from the chosen value of $\beta$ and its effect on the highest order derivative of the acceleration.
The inverse of $\underline{A}$ can be calculated from the LU decomposition of $\underline{A}$, $\underline{A}^{-1}=U^{-1}L^{-1}$. 
The $i,j$-th components of the inverse of the $L$ and $U$ matrices that generate the Pascal Matrix are given by
\begin{equation}
L^{-1}_{ij}=(-1)^{i+j}\left(\begin{smallmatrix} i\\j \end{smallmatrix}\right)
\end{equation}
and
\begin{equation}
U^{-1}_{ij}=
\left\{
\begin{array}{ll}
      \frac{(-1)^{i+j}}{2^{2j-i}}\frac{i}{j}\left(\begin{smallmatrix} 2j-i-1\\j-i \end{smallmatrix}\right)       & 1 \leq i \leq j \\
      1 & i=j=0\\
      0 & \rm otherwise \\
\end{array}
\right
.
\end{equation}
\citep{2016RIKEN.K.N}.
Thus, $\delta \textbf{x}$ corresponds to the $i=j=p+1$ term in $\underline{A}^{-1}$, such that 
\begin{equation}
\delta \textbf{x} = \frac{\beta-1}{2p+3}\frac{(-1)^{4p}}{2^{p+1}}\left(\begin{smallmatrix} p+1\\p+1 \end{smallmatrix}\right)\textbf{F}^\pm_{p+1}
= \frac{\beta-1}{2p+3}\frac{1}{2^{p+1}}\textbf{F}^\pm_{p+1}.
\end{equation}

For a time-symmetric integrator with time-symmetric time steps, the errors in the magnitude of the eccentricity vector of a Keplerian orbit should cancel over one period. However, errors may accumulate in the components of the eccentricity vector such that orbits precess artificially over time. Thus, we determine the value of $\beta$ such that the contributions of the leading truncation error terms in \textbf{x} and \textbf{v} lead to errors in components of the eccentricity vector ($e_i$) that integrate to $0$ over the course of an orbit, where
\begin{equation}
\delta e_i = \frac{\partial e_i}{\partial \textbf{x}}\cdot\delta \textbf{x}+\frac{\partial e_i}{\partial \textbf{v}}\cdot\delta \textbf{v}.
\end{equation}
This leading term should cancel over a Keplerian period, so 
\begin{equation}
\int_0^{T} \delta e_i ~dt = 0 = \int_0^{T} \left( \frac{\partial e_i}{\partial \textbf{x}}\cdot\delta \textbf{x}+\frac{\partial e_i}{\partial \textbf{v}}\cdot\delta \textbf{v}\right)dt,
\end{equation}
where $T$ is the Keplerian period. 

Then, recalling that $\textbf{F}_n$ has $2p$ derivatives of acceleration for $\delta \textbf{v}$ but $2p-1$ derivatives of acceleration for $\delta \textbf{x}$, we see that $\beta$ can be determined using
\begin{equation}
\begin{split}
\frac{\beta-1}{(2p+3)2^{p+1}} \int_0^{T}\left( \frac{\partial e_i}{\partial \textbf{x}}\cdot\textbf{a}^{(2p-1)}  \right)dt -\\ \frac{-1}{(-2)^{p+1}}\frac{(2p+2)!!}{(2p+3)!!}\int_0^{T} \left( \frac{\partial e_i}{\partial \textbf{v}}\cdot\textbf{a}^{(2p)} \right)dt = 0
\end{split}
\end{equation}
or equivalently
\begin{equation}\label{eq:I/J}
\beta = 1 + \left(\frac{(2p+3)2^{p+1}}{(-2)^{p+1}}\frac{(2p+2)!!}{(2p+3)!!} \right)\frac{\int_0^{T} \left( \frac{\partial e_i}{\partial \textbf{v}}\cdot\textbf{a}^{(2p)} \right)dt}{\int_0^{T}\left( \frac{\partial e_i}{\partial \textbf{x}}\cdot\textbf{a}^{(2p-1)}  \right)dt},
\end{equation}
where $\textbf{a}$ is the acceleration. 
We use the following identity, for which we provide a proof in Appendix \ref{apndx:A}, 
\begin{equation}\label{eqn:ident}
\frac{\partial e_i}{\partial \textbf{x}}=-\frac{d}{d t}\frac{\partial e_i}{\partial \textbf{v}}.
\end{equation}
Applying this to the denominator of equation (\ref{eq:I/J}), we find that
\begin{equation}\label{eq:identitied}
\int_0^{T}\frac{\partial e_i}{\partial \textbf{x}}\cdot\textbf{a}^{(2p-1)}dt=-\int_0^{T}\textbf{a}^{(2p-1)}\cdot\frac{d}{d t}\frac{\partial e_i}{\partial \textbf{v}}dt.
\end{equation}
Integrating by parts, we find that the right side of equation (\ref{eq:identitied}) becomes
\begin{equation}
\left[-a^{(2p-1)}\cdot\frac{\partial e_i}{\partial \textbf{v}}\right]^T_0 + \int_0^{T}\frac{\partial e_i}{\partial \textbf{v}}\cdot\textbf{a}^{(2p)}dt.
\end{equation}
Because Keplerian orbits are periodic, the first term vanishes, leaving us with values of $\beta$ for Hermite integrators of order $2(p+1)$
\begin{equation}
\beta = 1 + \frac{(2p+3)2^{p+1}}{(-2)^{p+1}}\frac{(2p+2)!!}{(2p+3)!!} = 1 + \left(-1\right)^{p+1}\frac{(2p+2)!!}{(2p+1)!!}.
\end{equation}
Thus, for the 4th-order Hermite scheme, with $p=1$, the optimal value is $\beta=11/3,$ which results in the corrector expression for position
\begin{equation}
\textbf{x}_1-\textbf{x}_0=\frac{1}{2}(\textbf{v}_1+\textbf{v}_0)\Delta t-\frac{7}{60}(\textbf{a}_1-\textbf{a}_0)\Delta t^2+\frac{1}{60}(\dot{\textbf{a}}_1+\dot{\textbf{a}}_0)\Delta t^3
\end{equation}
which matches that derived in \citet{2004PASJ...56..861K} for their parameterization $\alpha=7/6$,
where the subscript $0$ indicates quantities at the beginning of a time step and subscript $1$ indicates quantities at the end of a timestep,

Considering the 6th- and 8th-order schemes, improved position correctors may be expressed in the following way, with $\beta=-11/5$ for $p=2$ and $\beta=163/35$ for $p=3$:
\begin{equation} \label{eqn: corr6}
\begin{split}
\textbf{x}_1-\textbf{x}_0=\frac{1}{2}(\textbf{v}_1+\textbf{v}_0)\Delta t-\frac{4}{35}(\textbf{a}_1-\textbf{a}_0)\Delta t^2+ \\
\frac{13}{840}(\dot{\textbf{a}}_1+\dot{\textbf{a}}_0)\Delta t^3 -\frac{1}{840}\left(\textbf{a}^{(2)}_1 - \textbf{a}^{(2)}_0\right)\Delta t^4
\end{split}
\end{equation}
for the 6th-order corrector and 
\begin{equation}\label{eqn: corr8}
\begin{split}
\textbf{x}_1-\textbf{x}_0=\frac{1}{2}(\textbf{v}_1+\textbf{v}_0)\Delta t-\frac{29}{252}(\textbf{a}_1-\textbf{a}_0)\Delta t^2+\frac{1}{63}(\dot{\textbf{a}}_1+\dot{\textbf{a}}_0)\Delta t^3 \\ 
-\frac{1}{720}\left(\textbf{a}^{(2)}_1 - \textbf{a}^{(2)}_0\right)\Delta t^4 +  \frac{1}{15120}\left(\textbf{a}^{(3)}_1 + \textbf{a}^{(3)}_0\right)\Delta t^5
\end{split}
\end{equation}
for the 8th-order corrector.
From the form of equations \ref{eqn: corr6} and \ref{eqn: corr8} it is clear that these schemes are both time-symmetric and implicit. 
In practice, these schemes are evaluated using Predict-Evaluate-Correct (PEC) iterations. If we iterate $n$ times, this is denoted a P(EC)$^n$ scheme. The method is formally implicit and time symmetric if iterated until convergence, and taking $n$ between 2 and 4 is usually sufficient, as demonstrated in Section \ref{sec:numer}.

\section{Numerical Tests}\label{sec:numer}
We explore a selection of numerical examples to confirm the validity and applicability of our derivations. We test these methods using a P(EC)$^n$ scheme, where $n=1$ results in a standard Hermite scheme \citep{1992PASJ...44..141M} and $n>1$ leads to increasingly time-symmetric and implicit schemes. As demonstrated later, we find that increasing the number of iterations can be computationally favourable, facilitating good behaviour even for large time steps. Because the 6th- and 8th-order schemes require knowledge of acceleration and its derivatives in order to predict the positions of particles at the next time step, care must be taken during the initialisation of the algorithm. While \citet{2008NewA...13..498N} 
initialised runs using a lower-order scheme and sufficiently small timestep, we explicitly calculate higher-order derivatives of acceleration for the initialisation procedure. 

During each predictor step, we use a standard Taylor series. Thus, for the 8th-order scheme, we calculate up to the 5th derivative of acceleration during initialisation. The benefit of this method is that it guarantees that our initialisation procedure does not introduce any errors associated with the initial time step, and that it facilitates simulations with a constant time step. For each test in this section, we use a gravitational softening parameter $\epsilon$, where we set $\epsilon = 10^{-8}$ for our 2-body tests and $\epsilon=10^{-6}$ for our N-body tests.
For clarity, this means that we model the acceleration on particle $i$ due to interaction with particle $j$
as $ \textbf{a}_{ij}=Gm_j\textbf{r}_{ij}( r_{ij}^2 + \epsilon^2 )^{-3/2}$,
where $\textbf{r}_{ij}$ is the displacement vector between particles $i$ and $j$ and $r_{ij}$ is the magnitude of $\textbf{r}_{ij}$.

\subsection{Keplerian Orbits}
\begin{figure}
\includegraphics[width=\columnwidth]{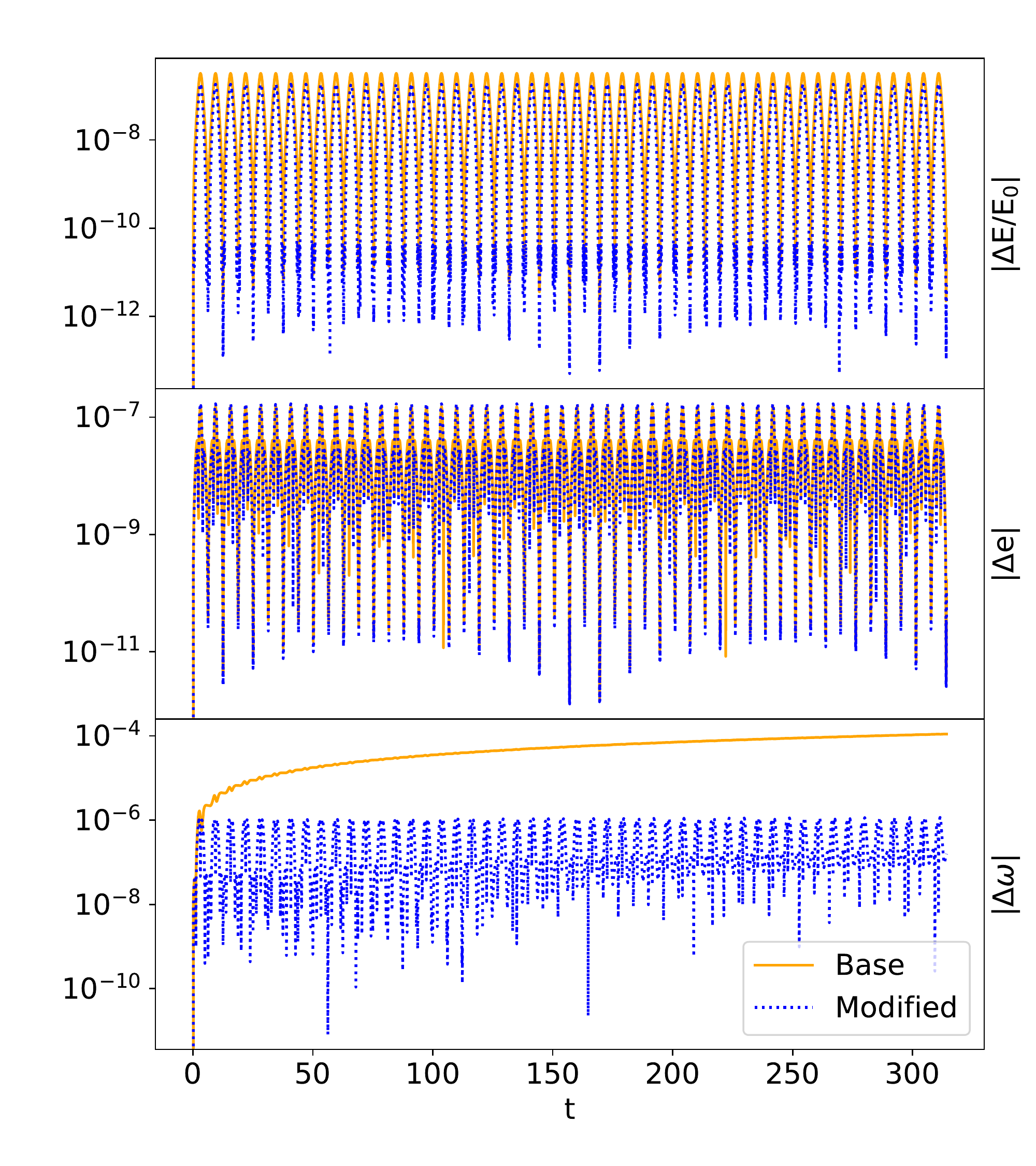}
\caption{Energy and selected orbital elements for a Keplerian orbit with $a=1.0$ and $e=0.1$ using the 4th-order Hermite algorithm. Both simulations used $n=3$ and $\Delta t=2^{-4}$. The orange solid line corresponds to the common Hermite scheme of \citet{1992PASJ...44..141M} and the blue dotted line corresponds to the modified algorithm of \citet{2004PASJ...56..861K} for which an alternative derivation was given in this work. The basic algorithm exhibits secular growth in argument of periapsis, whereas the modified algorithm does not.}
\label{fig: elem4}
\end{figure}

We begin by testing our schemes for simple Keplerian orbits. The 4th-order scheme has already been derived and tested by \citet{2004PASJ...56..861K}. We wish to verify that our implementation behaves as expected and provide a baseline for comparison with other methods. We consider a $10^{-3}~M_\odot$ planet orbiting a $1~M_\odot$ star, with a semi-major axis $a=1.0$ and eccentricity $e=0.1$. The system is evolved until time $t=100\pi$ with a time step of $\Delta t = 2^{-4}$ using $n=3$. Figure \ref{fig: elem4} plots the normalised energy error $\Delta E/E_0\equiv (E-E_0)/E_0$, along with errors in the eccentricity $\Delta e \equiv (e-e_0)$ and argument of periapsis $\Delta \omega \equiv (\omega-\omega_0$)/(1 \rm{rad}). As expected, both 
time-symmetric algorithms accrue no secular errors in energy or eccentricity. However, the basic algorithm undergoes secular growth in the argument of periapsis, whereas the modified algorithm accrues minimal secular errors. 
Thus, we have verified our implementation of the 4th-order time symmetric algorithm. We note that \citet{2004PASJ...56..861K} also demonstrated that the modified algorithm evolves the argument of periapsis more accurately for systems of multiple planets.

The 6th-order Hermite scheme was initially presented in \citet{2008NewA...13..498N}, where the authors noted that appropriate treatment of the highest-order terms would improve the behaviour of the integrator for Keplerian orbits, but neglected such treatment because it would not improve the order of the time integration. We present such a treatment here, and show that it can produce smaller energy errors in addition to reducing secular errors in the argument of periapsis. We contrast the corrector given by equation (\ref{eqn: corr6}) with the one presented in \citet{2008NewA...13..498N}. 

\begin{figure}
\includegraphics[width=\columnwidth]{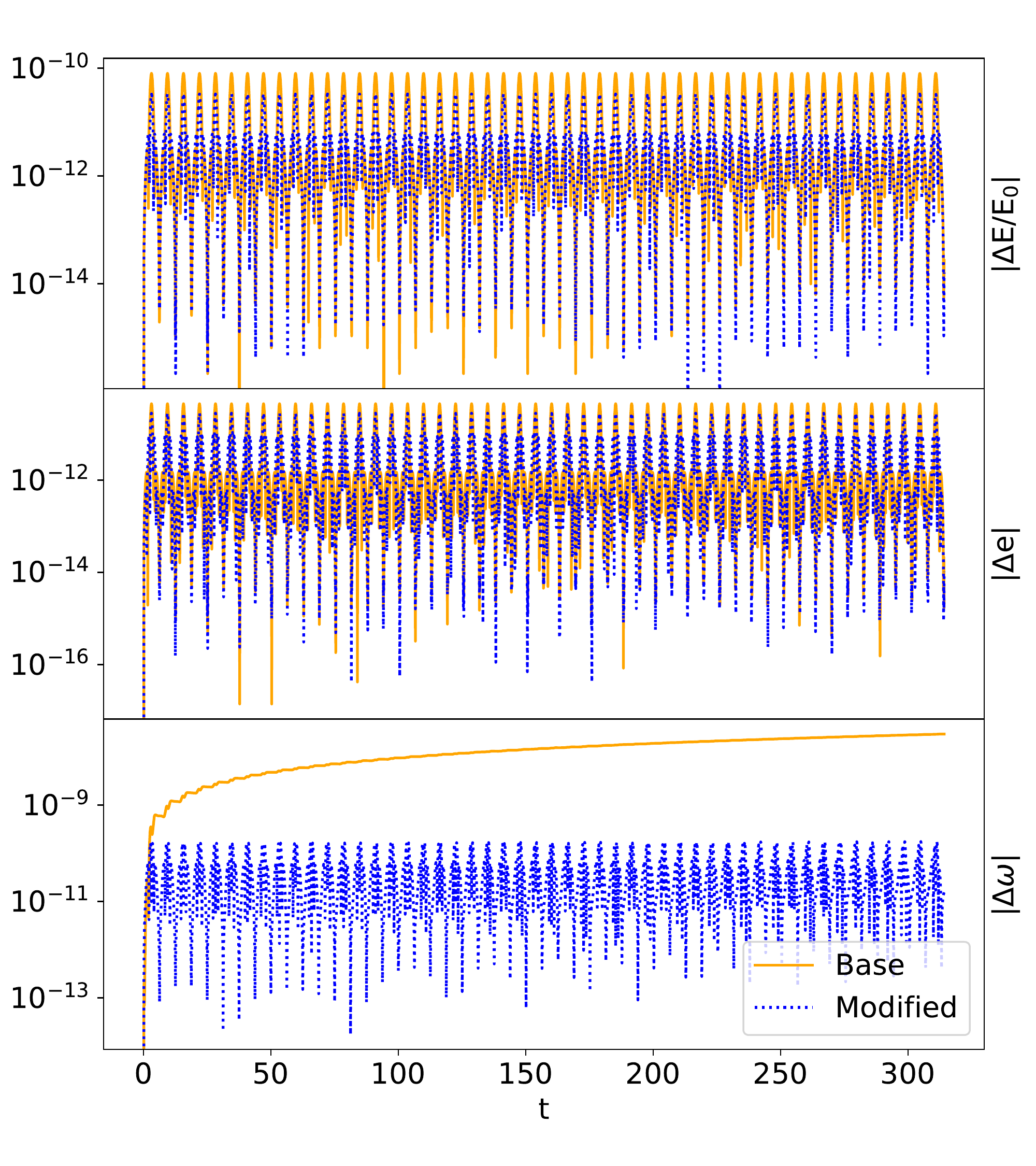}
\caption{Energy and selected orbital elements for a Keplerian orbit with $a=1.0$ and $e=0.1$ using 6th-order Hermite integrators with $n=3$ and $\Delta t=2^{-4}$. Results using the integrator presented in \citet{2008NewA...13..498N} are plotted in orange, while results using the modified algorithm derived here are plotted with a dotted blue line.}
\label{fig: elem6}
\end{figure}

We use the same initial conditions and simulation parameters as when testing the 4th-order scheme for Keplerian orbits. The results are presented in Figure \ref{fig: elem6}. Because these results and those in Figure \ref{fig: elem4} use the same time step, the advantages of the higher-order algorithm are clear: calculating the 2nd derivative of acceleration increases the number of floating point operations by a factor $\sim 1.6$. However, the energy error in Figure \ref{fig: elem6} is reduced by about three orders of magnitude compared to that in Figure \ref{fig: elem4}, whereas decreasing the time step of a 4th-order algorithm by a factor of 2 would reduce the energy error by a factor of $\sim 16$. Additionally, the modified algorithm described by equation (\ref{eqn: corr6}) exhibits the preferable qualities of the 4th-order algorithms presented in \citet{2004PASJ...56..861K}, specifically that the energy, eccentricity, and argument of periapsis do not exhibit significant secular growth.

\begin{figure}
\includegraphics[width=\columnwidth]{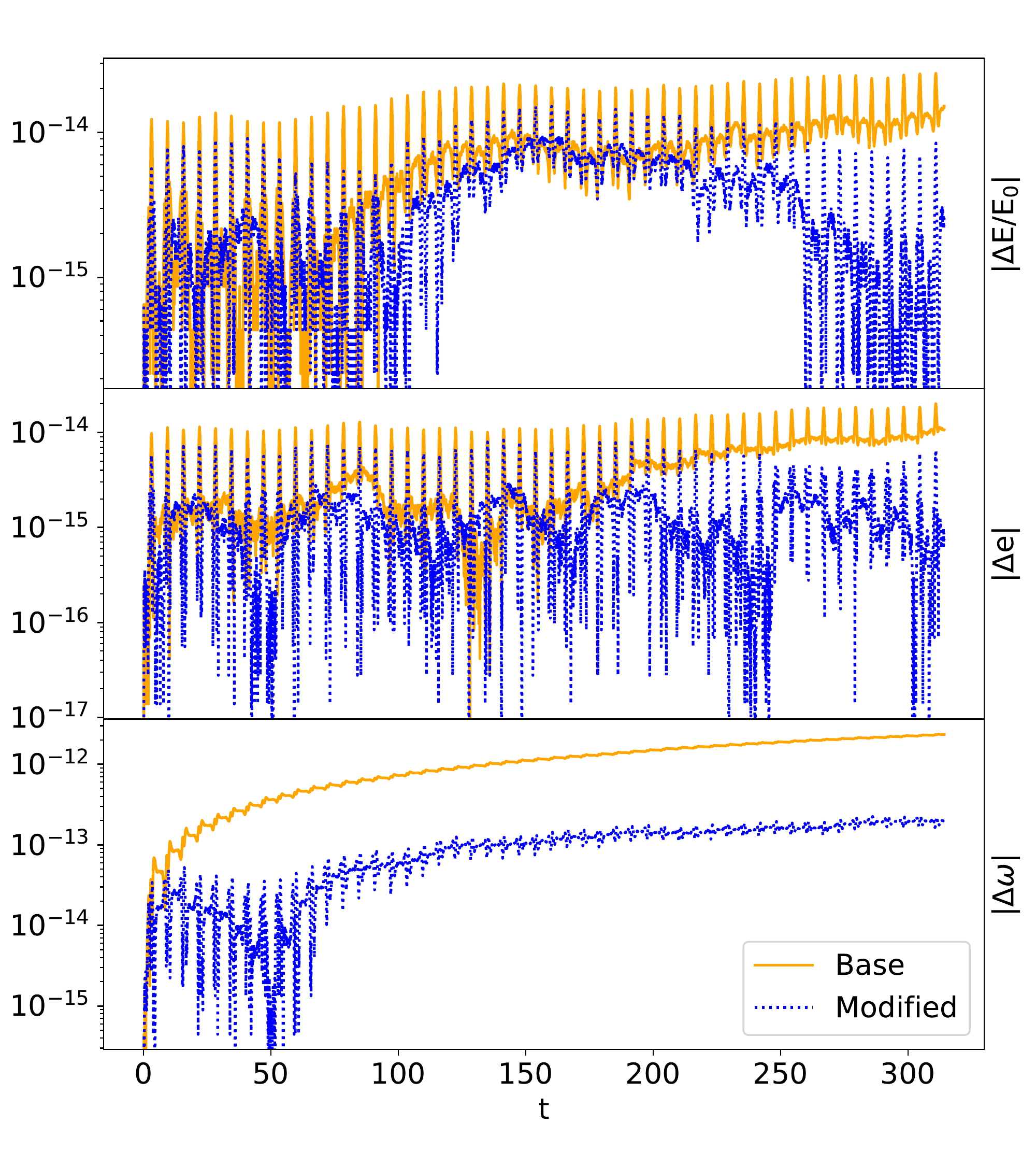}
\caption{Energy and selected orbital elements for a Keplerian orbit with $a=1.0$ and $e=0.1$ using 8th-order Hermite algorithms with $n=3$ and $\Delta t=2^{-4}$. The orange solid line plots results from the Hermite algorithm presented in \citet{2008NewA...13..498N}, while the blue dotted line plots results using the integrator derived in this work.
Even though rounding error dominates truncation error for both algorithms, the standard algorithm exhibits much larger secular growth in the argument of periapsis. Note that in this case, most of the growth of the argument of periapsis for the modified algorithm is due to the gravitational softening $\epsilon\neq0$, which causes precession because the force is no longer an inverse-square law.}
\label{fig: elem8}
\end{figure}

The 8th-order Hermite scheme was also initially presented in \citet{2008NewA...13..498N}. Here, we present the results of improving the truncation error when calculating changes in position, contrasting the corrector given by equation (\ref{eqn: corr8}) with the one presented in \citet{2008NewA...13..498N}. In figure \ref{fig: elem8} we present results using the same test as for the 4th- and 6th-order integrators. In this case, a time step of $\Delta t=2^{-4}$ is sufficiently small for the integration error to be dominated by round-off errors. However, even when both the errors in energy and eccentricity are dominated by round-off, the argument of periapsis still accrues secular errors using the corrector presented in \citet{2008NewA...13..498N}. Additionally, the 8th-order integrator has energy errors $2-3$ orders of magnitude smaller than the 6th-order integrator. Explicitly calculating the third derivative of acceleration increases the number of floating point operations by a factor of $\sim1.5$, whereas using half as small a time step for the 6th-order algorithm would only reduce the error by a factor of $\sim 64$. 

\begin{figure}
\includegraphics[width=\columnwidth]{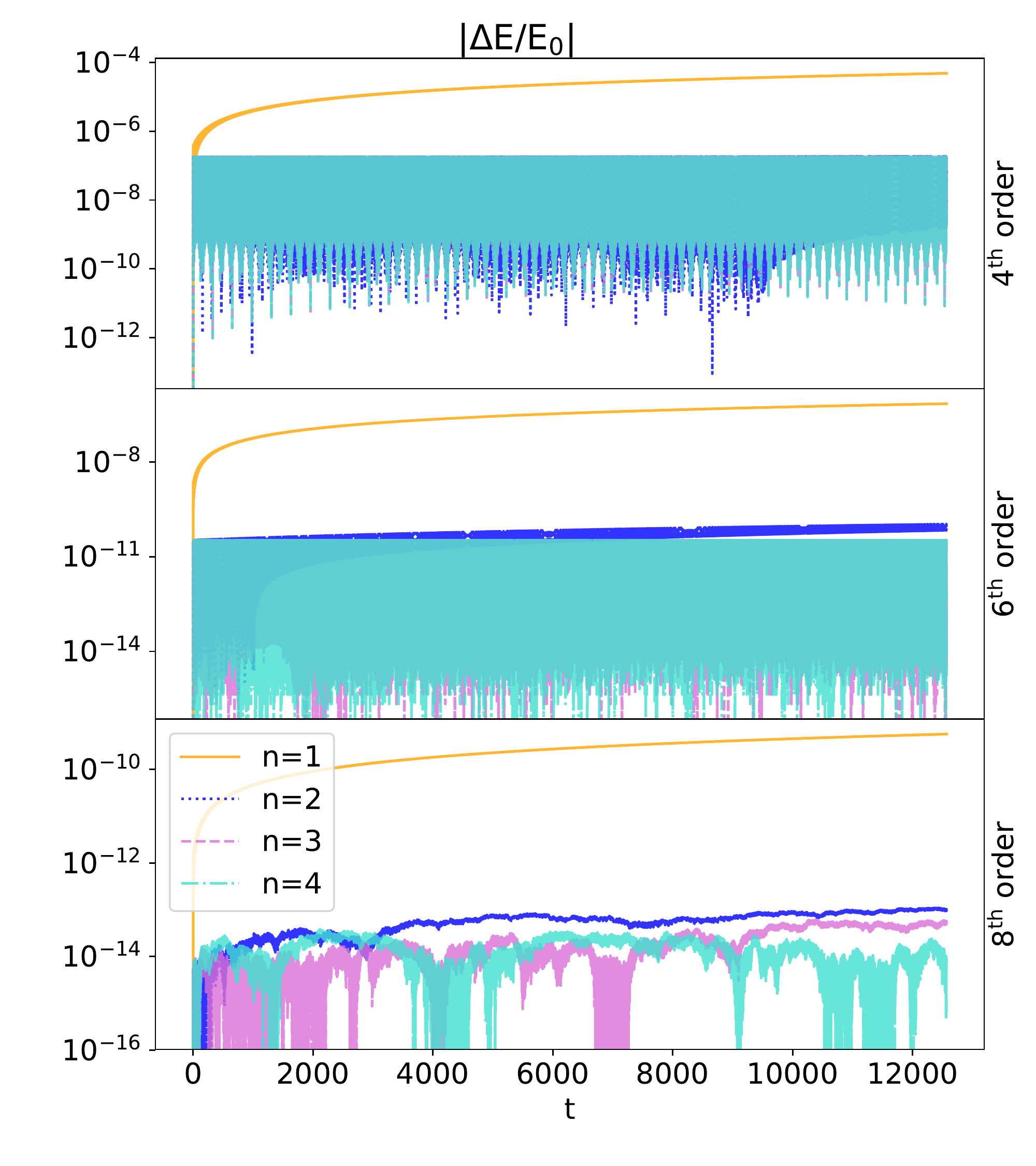}
\caption{Energy errors for the modified Hermite methods discussed in this paper. The top panel plots energy error using the 4th-order modified scheme. The middle panel plots energy error using the 6th-order modified scheme, and the bottom panel plots the energy error using the 8th-order scheme. Orange solid lines denote the results using $n=1$. Dark blue dotted lines indicate results using $n=2$. Dashed purple lines indicate results using $n=3$, and dash-dotted turquoise lines indicate results using $n=4$. Results using $n=3$ and $n=4$ are largely indistinguishable. Integrations using $n=2$ exhibit slow growth of secular errors, 
but still achieve orders of magnitude smaller errors than those obtained using $n=1$}
\label{fig: lon}
\end{figure}

In all of the tests so far, we have used $n=3$ evaluation and correction iterations. It is informative to examine the behaviour of these integrators for different values of $n$. We present the results of these tests in figure \ref{fig: lon}, which investigates energy errors evolving the same system as the previous tests, but this time over the course of 2000 orbital periods. It is clear that $n=2$ is radically preferred over $n=1$ in each case. By careful examination, one can see that secular errors in energy grow in the $n=2$ case, whereas the $n=3$ and $n=4$ cases behave nearly identically. These results are similar to those found in \citet{2004PASJ...56..861K} for the 4th-order method, and it is reassuring to see that the higher-order methods behave similarly. 

\subsection{Few-body simulations}
Over the last two decades, numerous systems containing multiple exoplanets have been discovered. One may wish to study the orbital stability of these systems, or to fit orbital models to transit or radial velocity data. 
The 4th-order method of \citet{2004PASJ...56..861K} 
has been utilised for studies such as these in the past, for example by \citet{2014ApJS..210...11N}. 
Because the cost of running N-body simulations can often dominate these studies, it is worthwhile to see how higher-order methods can benefit the study of few-body systems. 

As a test problem, we consider the orbits of WASP-47b, WASP-47d, and WASP-47e about the star WASP-47 \citep{2015ApJ...812L..18B} over the course of 5 years. This system consists of a hot Jupiter, a Neptune analogue outside the hot Jupiter orbit, and a Super-Earth interior to the hot Jupiter orbit. 
The star in the system has a mass of $1.04~M_\odot$. Planet properties are given in order from nearest to furthest from the star, with masses (in terms of Jupiter's mass, $M_J$) $0.0251~M_J$, $1.1424~M_J$, $0.0412~M_J$; semi-major axes $0.017~\rm{AU}$, $0.052~\rm{AU}$, and $0.085~\rm{AU}$; and eccentricities $0.03$, $0.0$, and $0.01$. The shortest orbital period is 0.79 days, and the initial condition was based on data from the Mikulski Archive for Space Telescopes\footnote{\url{https://exo.mast.stsci.edu/}.}.

Two important orbital elements for comparing N-body simulations with radial velocity and transit data from exoplanets are the argument of periapsis and the time of periapsis. Since the modified Hermite algorithms presented in this paper accrue minimal secular errors in the argument of periapsis, the other relevant metric by which to compare them is error in the time of periapsis. Because error in the time of periapsis is proportional to the integral of error in mean motion, which is proportional to error in energy, we compare the maximum energy errors of the 4th, 6th, and 8th-order schemes for the aforementioned exoplanet system. For the 8th-order scheme, we also demonstrate use of compensated summation during the time integration to reduce rounding errors \citep{10.1145/363707.363723}.   

\begin{figure}
\includegraphics[width=\columnwidth]{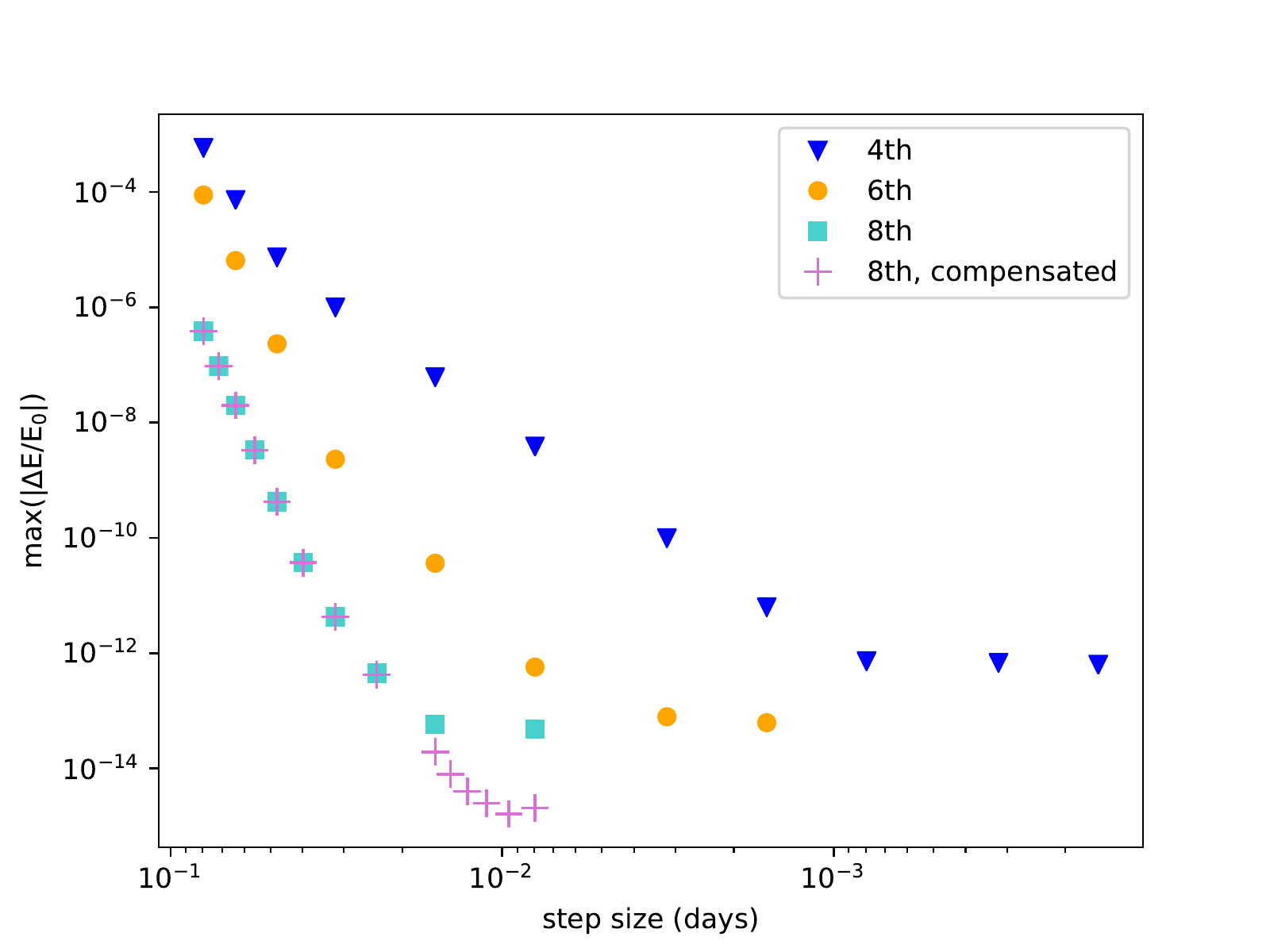}
\caption{Maximum energy errors for the few-body test problem integrated over 5 years. Blue triangles indicate results for the 4th-order algorithm, orange circles indicate results for the 6th-order algorithm, and turquoise squares indicate results for the 8th-order algorithm. Purple $+$ symbols indicate results using the 8th-order algorithm and compensated summation for the time integration. Each simulation was run using $n=3$ iterations.}
\label{fig:exo}
\end{figure}

When interpreting the relative costs and benefits of different schemes, it is important to recall the additional cost in floating point operations of each scheme. The 6th-order scheme requires about 1.6 times as many floating point operations as the 4th-order scheme, and the 8th-order scheme requires about 1.5 times as many floating point operations as the 6th-order scheme \citep{2008NewA...13..498N}. The 4th-order scheme reaches a minimum energy error on the order of $10^{-12}$, while the 6th-order scheme is able to achieve similar energy errors using roughly nine times larger a step size. Assuming one would like to minimise errors in double precision, the 6th-order scheme is roughly four times more efficient. 
The 8th-order scheme is marginally more efficient than the 6th-order scheme for achieving rounding-dominated errors.
If one can tolerate larger errors, the 4th-order scheme may be preferable. However, with higher-order schemes, energy errors on the order of $10^{-15}$ can be achieved economically. 

\subsection{N-body simulations}
One might be concerned with applying higher-order methods to collisional problems because of their reliance on the smoothness of higher-order derivatives of acceleration. Since this concern is more pronounced for the 8th-order algorithm presented here than for the 6th-order algorithm, we test the 8th-order algorithm in an N-body problem.

In order to model collisional systems, either a very small constant time step or a variable time step is necessary. We choose the latter, adopting the procedure described in \citet{1995ApJ...443L..93H}. For our time step criterion, we chose the minimum two-body Keplerian period, multiplied by a factor $\eta$.
Thus, we express our timestep criterion $H(\textbf{x}, \textbf{v}...)$ as 
\begin{equation}\label{eq:dt}
H(\textbf{x}) = \eta ~\min_{i\neq j}\left(\frac{(r^2_{ij}+\epsilon^2)^{3/2}}{m_i+m_j}\right)^{1/2}, 
\end{equation}
where $\eta$ controls the accuracy of the scheme. This simple criterion is clearly sub-optimal in the sense that it does not use all of the available information about acceleration and its derivatives. However, this criterion serves to demonstrate the robustness of the method. The time step is chosen implicitly as the Hermite method iterates, using $\Delta t = 0.5(H_+ + H_-)$ where $H_+$ and $H_-$ are given, respectively, by equation (\ref{eq:dt}) evaluated at the beginning and end of the timestep. After a few iterations, such a time step is both variable and symmetric in time. 

Because the development of these methods was inspired by dynamics in discs, we would like to test their performance in such scenarios. We choose to simulate a disc of 100 objects with masses $10^{-5} M_\odot$ around a central solar-mass object. We draw the initial eccentricities and inclinations from Rayleigh distributions with scales $\sigma_i = 2\sigma_e = 0.12$. Initial semi-major axes are chosen with logarithmic spacing between 0.1 and 10 AU. We simulated these systems until time $t=100\pi$ using softening parameter $\epsilon=10^{-6}$. We explore various choices of $n$ and $\eta$. We tabulate the median energy error $\delta E = |E - E_0|/|E_0|$ over the last 
unit of simulation time in table \ref{tab:enres}. 

\begin{table}
\begin{centering}
\begin{tabular}{cccccc}
    $\delta E$         & $\eta=0.08$ & $\eta=0.04$ & $\eta=0.02$ & $\eta= 0.01$ & $\eta=0.005$\\ \hline
 $n$ = 3 & $2.4\mathsmaller{\times}10^{-4}$& $1.0\mathsmaller{\times}10^{-5}$ & $3.9\mathsmaller{\times}10^{-7}$ & $6.5\mathsmaller{\times}10^{-9}$ & $3.2\mathsmaller{\times}10^{-10}$\\
 $n = 4$ & $4.2\mathsmaller{\times}10^{-6}$& $3.4\mathsmaller{\times}10^{-8}$ & $7.8\mathsmaller{\times}10^{-10}$ & $3.0\mathsmaller{\times}10^{-13}$ & $2.3\mathsmaller{\times}10^{-13}$ \\
\hline
\end{tabular}
\end{centering}
\caption{Energy errors for our N-body test, for various time step parameters, using the 8th-order modified Hermite algorithm.}
\label{tab:enres}
\end{table}

For comparison, we ran one simulation with the same initial conditions and time step criterion, but using the 4th-order Hermite integrator,
$n=4$, and $\eta = 0.08$. We find that the median energy error over the last time unit is $\delta E = 1.9 \times 10^{-4}$. Note that reducing $\eta$ by a factor of $~2.6$ when using the 4th-order method could yield comparable energy error to the 8th-order integrator using $\eta=0.08$. However, the 8th-order algorithm only requires about $2.4$ times the floating point operations of the 4th-order algorithm each time step. Thus, it is clear that the 4th-order algorithm is preferable for low-accuracy simulations, but for high-accuracy simulations higher-order methods are clearly preferable. 

\section{Discussion}
The schemes presented here, and the higher-order variations that follow from these formulae, are well-suited to study systems governed by a central potential, even when close encounter frequently occur. These methods are therefore highly applicable to the study of planetesimal discs \citep{2010ApJ...714L..21K} and stellar discs around black holes \citep{{2003ApJ...590L..33L},{2020MNRAS.493.3732D}}. 
Additionally, these schemes offer benefits for cluster simulations and gravitational scattering experiments where binaries can be evolved with a constant time step for a non-negligible portion of their evolution. The improved treatment of binaries and trivial cost of the modified Hermite integrators compared to the ones presented in \citet{2008NewA...13..498N} may make them worth implementing in cluster simulations even though the overall energy conservation would not change significantly. 

Time-symmetric Hermite schemes can be implemented using a shared time step to maintain the time-symmetry of all particles, or using a block time step \citep{2006NewA...12..124M}. Considering block time stepping schemes, the higher-order methods presented here have a number of computational advantages over the 4th-order time symmetric scheme: not only does accuracy increase faster than computational cost (e.g. Figure \ref{fig:exo}), but the number of particles integrated in one block step is larger for higher-order schemes \citep{2008NewA...13..498N}. Additionally, the extra work in higher-order schemes is highly parallel, a benefit of Hermite schemes in general over other high-order schemes, making these schemes attractive for implementation on highly parallel architectures. We also note that \cite{2003PASJ...55.1163M} find good double-precision performance calculating higher-order terms using lower precision, which could further increase performance using accelerators such as GPUs.

High-order time-symmetric Hermite schemes also hold promise for extended-precision integrations. Given how the 8th-order scheme presented here is able to readily push the limits of double precision arithmetic, higher-order schemes could integrate planetary systems using a reasonably large time steps and still reach the limits of quadruple precision. High-order Hermite methods have the additional benefit that they can be made time symmetric, greatly reducing the energy error and phase error of the schemes in long-duration simulations. We also note that the 4th-order Hermite scheme has had difficulty integrating systems involving black hole binaries and a star cluster about a black hole: specifically, \citet{2016ApJ...828...77V} found that NBODY6 \citep{2012MNRAS.424..545N} had difficulty evolving the close approaches of stars to the black hole due to the time step being pushed towards machine precision. Such problems could be avoided with higher-order methods, since larger time steps can be used to achieve sufficient accuracy. Additionally, high-order methods combined with higher-precision arithmetic hold promise for evolving systems with extreme mass ratios, such as the nine orders of magnitude found between stars and SMBHs. 

\section{Conclusions}
We have presented a derivation of Hermite schemes of arbitrary order, along with a generic formula to optimise the behaviour of the integrators for Keplerian orbits. We have verified our derivations through numerical examples ranging from 2-body to N-body systems, using constant and variable time-symmetric time steps. Our 2-body tests demonstrate severe reduction of secular errors in the argument of periapsis and small decreases in energy and eccentricity errors for Keplerian orbits. The integrators presented here trivially extend to systems with velocity-dependent forces, and the derivations presented here for numerically preserving the argument of periapsis can be modified to preserve conserved quantities in other systems. 

The schemes presented here achieve extremely small errors that are bounded in time more economically than the common 4th-order scheme. Given the arbitrarily high-order nature of the derivations, these schemes can easily be extended to reach arbitrary accuracy. These schemes hold promise for the economical evolution of few-body systems, binaries within clusters, and other centrally-dominated potentials. 

\section*{Acknowledgements}
The author is thankful to Cole Miller, Derek Richardson, and Zeeve Rogoszinski for advice on preparation of the manuscript. The author also thanks Jun Makino and Keigo Nitadori for useful discussions, and comments on a draft of this work. 


\bibliographystyle{mnras}
\bibliography{references.bib}



\appendix
\section{Derivatives of Eccentricity}\label{apndx:A}
Since equation (\ref{eqn:ident}) is not obvious, we demonstrate it here. We begin by multiplying the eccentricity vector $\textbf{e}$ by the standard gravitational parameter $\mu$ to form the Laplace-Runge-Lenz vector ($\bar{\textbf{e}})$, a choice that makes no difference in the result but simplifies notation:
\begin{equation}
\bar{\textbf{e}}\equiv\mu \textbf{e}=\textbf{v}\times\left(\textbf{r}\times\textbf{v}\right)-\mu\frac{\textbf{r}}{|\textbf{r}|}
=\left(|\textbf{v}|^2-\frac{\mu}{\textbf{|r|}}\right)\textbf{r}-\left(\textbf{r}\cdot\textbf{v}\right)\textbf{v}
\end{equation}
and a component of $\bar{\textbf{e}}$ is 
\begin{equation}
\bar{e}_i=\left(v_m v^m-\frac{\mu}{\sqrt{r_k r^k}}\right)r_i - \left(r_l v^l\right)v_i.
\end{equation}
Then, because $\partial v_i/\partial r_j = \partial r_i/\partial v_j = 0$ and
$\partial r_i/\partial r_j = \partial v_i/\partial v_j = \delta_{ij}$ where $\delta_{ij}$ is the Kronecker delta, 
\begin{equation}\label{eqn:dedr}
\begin{split}
\frac{\partial \bar{e}_i}{\partial{r_j}}=-\mu\left(\frac{\delta_{ij}}{\sqrt{r_k r^k}} - \frac{r_i\delta_{kj} r^k}{\sqrt{r_k r^k}^3} \right)-v_i\left(\delta_{lj} v^l\right)
+\delta_{ij}v_m v^m\\
=\frac{\mu}{\left(r_k r^k\right)^{3/2}}\left(r_i r_j  - \delta_{ij}r_l r^l\right)-v_i v_j+\delta_{ij} v_m v^m
\end{split}
\end{equation}
\begin{equation}\label{eqn:edv}
\frac{\partial \bar{e}_i}{\partial v_j} = 2\delta_{mj} v^m r_i- r_lv^l\delta_{ij}-\delta_{lj}r_l v_i=2v_jr_i-r_jv_i-\delta_{ij}r_lv^l.
\end{equation}
Then, taking the time derivative of \ref{eqn:edv},
\begin{equation}\label{eqn:ddtdedv}
\begin{split}
\frac{d}{dt}\frac{\partial \bar{e}_i}{\partial v_j}=2a_jr_j+2v_jv_i-v_jv_i-r_ja_i-\delta_{ij}\left(v_lv^l + r_la^l \right)\\
=2a_jr_i-a_ir_j+v_iv_j-\delta_{ij}\left(v_lv^l+r_la^l\right).
\end{split}
\end{equation}
Then, recall that 
\begin{equation}
a_i = -\mu\frac{r_i}{\left(r_lr^l\right)^{3/2}},
\end{equation}
so that equation (\ref{eqn:ddtdedv}) becomes
\begin{equation}\label{eqn:dteq}
\frac{d}{dt}\frac{\partial \bar{e}_i}{\partial v_j}=-\frac{\mu}{\left(r_kr^k\right)^{3/2}}\left(r_ir_j - \delta_{ij}r_lr^l\right)+v_iv_j - \delta_{ij}v_mv^m.
\end{equation}
Then, it is clear by inspection of equations (\ref{eqn:dteq}) and (\ref{eqn:dedr}) that 
\begin{equation}
\frac{\partial e_i}{\partial{r_j}}=-\frac{d}{dt}\frac{\partial e_i}{\partial v_j}.
\end{equation}


\bsp	
\label{lastpage}
\end{document}